\documentclass[12pt]{article}
\input{epsf}
\begin{document}
%
%
\setcounter{page}{0}
\thispagestyle{empty}
\vspace{-2cm}
\hfill{RUB-TPII-04/99}

\begin{center}
%
%
\begin{Large}{\bf The electroproduction of the $\Delta(1232)$ \\
                  in the chiral quark-soliton model} 
\end{Large} \\
\vspace{.5cm}
%
%
{\bf 
     A. Silva$^{{\rm a,d,}}$\footnote{Email: ajsilva@tp2.ruhr-uni-bochum.de},
     D. Urbano$^{\rm a,b,}$\footnote{Email: dianau@tp2.ruhr-uni-bochum.de},
     T. Watabe$^{\rm c,}$\footnote{Email: watabe@rcnp.osaka-u.ac.jp},
     M. Fiolhais$^{\rm d,}$\footnote{Email: tmanuel@teor.fis.uc.pt} 
 and K. Goeke$^{\rm a,}$\footnote{Email: Klaus.Goeke@tp2.ruhr-uni-bochum.de}   
} 

\vspace{.5cm}
%
%
{\small{\em 
$^{\rm a}$ Institut f\"{u}r Theoretische Physik II, Ruhr-Universit\"at
           Bochum,  D-44780 Bochum, Germany \\
$^{\rm b}$ Faculdade de Engenharia da Universidade do Porto, Rua dos
           Bragas, P-4000 Porto, Portugal\\
$^{\rm c}$ Research Center for Nuclear Physics (RCNP), Osaka
           University, Osaka 567, Japan \\
$^{\rm d}$ Departamento de F\'\i sica and Centro de F\'\i sica 
           Computacional,\\ Universidade de Coimbra, P-3000, Portugal\\ }}
\end{center}
\vspace{1cm}
%
%
\begin{abstract}
We calculate the ratios E2/M1 and C2/M1 for the 
electroproduction of the  $\Delta (1232)$ in the 
region of photon virtuality $0\!<\!\! -q^{2}\!<\! 1\,\,\mbox{GeV}^2 $. 
The magnetic dipole amplitude M1 is also presented. The theory used is the
chiral  quark-soliton model, which is based on the instanton vacuum of the
QCD.  The calculations are performed in flavours SU(2) and SU(3) taking
rotational ($1/N_c$) corrections into account. The results for the ratios
agree qualitatively with the available data, although the magnitude of both
ratios seems to underestimate the latest experimental results.
\end{abstract}
\vspace{1cm}
\noindent PACS numbers: 12.39Fe; 13.40-f; 14.20-c  \\
          Keywords: Chiral quark-soliton model; Delta electroproduction
\vfill
\newpage
%
%
%
%
%
%

\section{Introduction}

Electromagnetic processes have always played an important part
in understanding the structure of the nucleon. Among these processes,
the electroproduction of the $\Delta(1232)$ deserves a special
place. In fact, from the viewpoint  of the simplest quark model, 
the process $\gamma^{\ast }N\rightarrow \Delta $
is expected to proceed through the spin flip of one of the quarks,
which implies a non-vanishing magnetic dipole transition
and vanishing quadrupole transition amplitudes. 
This contradicts the experimental observations of non-vanishing 
quadupole amplitudes, although small when
compared to the magnetic dipole amplitude.

The fact that the first available data~\cite{old_data1,old_data2} 
for the ratios
of electric and coulomb quadrupole amplitudes to the magnetic dipole 
amplitude, E2/M1 and C2/M1 respectively, did not establish a definite
picture for these ratios, motivated in recent years 
a renewed interest in measurements of such
electroproduction amplitudes, mainly 
due to favourable conditions offered by new experimental facilities 
like MAMI (Mainz), ELSA-ELAN (Bonn), 
LEGS (BNL), Bates (MIT) and the Jefferson Laboratory (Newport News). 
Results coming from these recent experimental activities 
are expected to allow for
a great improvement in the experimental knowledge of the ratios E2/M1 and
C2/M1, as \cite{photon_point,new_data,Got99} let anticipate.

The difference between the quark model predictions and experiment
caused intensive discussions on the structure of the nucleon, 
the delta and the transition density. This process has been
studied in the context of many models of baryon structure, which made 
an overall correct description of these ratios to become a 
constraint in model-making and refining.
In the context of the quark model, 
to accomodate these noticeable amplitude ratios, charge
deformations due to {\it d}-state admixtures in non-relativistic 
and relativized models have been invoked for an 
explanation~\cite{quark_model,Cap90}. Other explanations can nevertheless be
found: it has recently been estimated in photoproduction~\cite{Buc98}, 
that the proper consideration of two-body exchange currents alone 
can give raise to non-vanishing quadrupole amplitudes.
Other quark model formulations, like
a quark model constructed in the infinite momentum 
frame~\cite{Azn93}, a light-front constituent quark 
model~\cite{Car96} and a three-body force model~\cite{Aie96},
have also been used in addressing the $\Delta(1232)$ electroproduction
amplitudes and/or form factors.

Calculations regarding the electroproduction of the $\Delta(1232)$
were also carried out in the framework of models fully based on chiral
symmetry such as the 
$\sigma $- and chromodielectric models~\cite{Fio96},  
the Skyrme model~\cite{Wal97}, the cloudy bag 
model~\cite{Lu97} and also in heavy baryon
chiral perturbation theory~\cite{Ber94}, with a recent
work~\cite{Gel98} extending it. 
Another class of models is based on effective lagrangian densities
containing, to different extents, ingredients like  Born terms, 
vector mesons and nucleon resonances (see \cite{Dre99} and references therein).

It is the aim of this paper to study the ratios E2/M1 and
C2/M1 and their momentum dependence for the 
electroprodution of the $\Delta (1232)$ in
the chiral quark-soliton model (CQSM)~\cite{Dia97}, 
rewiewed in~\cite{Chr96,Alk96}.
This model is based on the interaction of quarks with Goldstone bosons
resulting from the spontaneous breaking of chiral symmetry.
In the form used here, it can be regarded both as
a version of an effective model derived from the instanton liquid
model~\cite{Dia86} with constant constituent quark mass~\cite{Dia88}
and as a non-linear version of the semi-bosonized 
Nambu-Jona-Lasinio model~\cite{Nam61}. It allows for an
unified description of mesons and baryons, both in flavours SU(2) and
SU(3), with a small number of free parameters~\cite{Chr96}, i.e. three in
SU(2) and four in SU(3)).
 A successful description of static properties
of baryons,
 like mass splittings~\cite{Blo93,Wei92}, axial 
constants~\cite{Blo96}, magnetic moments~\cite{Kim96} 
and form factors~\cite{Chr95,Kim96a} has been achieved.

In this work we generalize the work 
of Ref.\cite{Wat95} from photoproduction  to electroproduction 
and consider also the flavour SU(3). The SU(2) calculation is done in
exactly the same way as described and applied to many observables
in Ref.~\cite{Chr96}. This includes the evaluation of rotational
$1/N_c$-corrections in order to improve axial and magnetic properties. For
the calculation in SU(3) we apply the symmetry conserving approximation
recently developed in Ref.\cite{Pra99}. 

The outline of the paper is as
follows: In sections $2$ and $3$
 we give a short account of 
the model, in the
vacuum and baryonic sectors respectively, 
together with references to 
 the
original literature.
 Section $4$ contains 
the application of the model to the
electroproduction of the $\Delta(1232)$,
whose results for the amplitude 
ratios E2/M1 and  C2/M1 are presented in section $5$ together with
their discussion. In Section $6$ we present a summary of the 
main points discussed and the conclusions.  

\section{The chiral quark-soliton model (CQSM)}

Thc Chiral Quark Soliton Model (CQSM) has been developed in
Refs.\cite{Dia86,Dia88} and is described in detail in
Refs.\cite{Dia97,Chr96,Alk96} with numerous successful applications to the
baryons of the octet and decuplet, in particualr the nucleon and its form
factors (for a review see \cite{Chr96}). The approach is (shortly) reviewed in
this section in order to make the present paper selfcontained. 
   
The CQSM is a 
field theoretical model based on the following quark-meson Lagrangian  
\begin{equation}   {\cal L}_{\rm CQSM}=\bar{\Psi}(
i\gamma^\mu\partial_\mu-m-MU^{\gamma_5})\Psi .
\label{lagrangian}
\end{equation}
 
In this expression,  $m=diag(\bar{m},\bar{m},m_s)$,
$\bar{m}=(m_u+m_d)/2$ is the current mass matrix of the quarks 
 neglecting
isospin breaking 
 and $M$ is the dynamical quark mass which results from
 the
spontaneous chiral symmetry breaking. 
 The field $\psi$ is the
constituent quark field and $U^{\gamma_5}$
 designates the chiral
Goldstone boson field 

\begin{equation}
U^{\gamma_5}=\frac{1+\gamma_5}{2}U +\frac{1-\gamma_5}{2}U^\dagger \, , 
\end{equation}
with $U$ given by\footnote{In the following, 
$U$ without suffix will be used when both SU(2) and SU(3)
flavours are meant.}
\begin{equation}  U_{{\rm SU(2)}}(\mbox{\boldmath $x$})
        =e^{i\mbox{\boldmath $\scriptstyle\tau$}\cdot \vec{\theta}
        (\mbox{\boldmath ${\scriptstyle x}$})}         
     \quad\quad\quad{\rm and}\quad\quad\quad
     U_{{\rm SU(3)}}(\mbox{\boldmath $x$})=\left(
      \begin{array}{cc} U_{\rm SU(2)} (\mbox{\boldmath $x$}) & 0 \\ 0 & 1
  \end{array}    \right) ,          \end{equation}
assuming the well established \cite{Witten} embedding of SU(2) 
in SU(3), which is known to ensure the restriction of the right 
hypercharge in such a way that the octet and
 decuplet are the 
lowest possible representations. 

The lagrangian (\ref{lagrangian}) corresponds to a non-renormalizable 
field theory. Therefore, the practical applications must be done within 
a certain regularization method, which becomes part of the model. 
As in most of the applications \cite{Chr96}, in this work we use 
the proper-time regularization (\ref{eq:reg}), since we prefer to work 
with the euclidean version of the lagrangian (\ref{lagrangian})
\begin{equation}
 {\cal L}=\psi^\dagger D(U)\psi \label{eq:lageuc}\end{equation}
where, choosing to work with hermitean euclidean Dirac matrices 
$\gamma^\dagger_\mu=\gamma_\mu$, the operator
\begin{equation}
D(U) = \gamma_4(\gamma_\mu\partial_\mu+m+MU^{\gamma_5})
     = \partial_4+h(U)+\gamma_4(m-\bar{m}\mathbf{1})
\end{equation}
includes the euclidean time derivative $\partial_4$ and the 
Dirac one-particle hamiltonian
\begin{equation} h(U) = \gamma_4(\gamma_k\partial_k
                      + \bar{m}\mathbf{1} + MU^{\gamma_5}).
\label{hamiltonian}   \end{equation}

In SU(3), due to the embedding, a projection onto strange, 
$S={\rm diag}(0,0,1)$, and non-strange, $T={\rm diag}(1,0,0)$,
subspaces gives:
\begin{equation}
D^{-1}(U)= D^{-1}(U_{\rm SU(2)})T+D^{-1}(U=1)S,
\label{eq:dsu3}\end{equation}
\begin{equation}
h(U)=h(U_{SU(2)})T+h(U=1)S,
\label{eq:hsu3}\end{equation}
with $U=1$ the vacuum configuration.

\subsection{Fixing the parameters}
Neglecting isospin breaking, two of the three parameters of the model in
SU(2) are the current quark mass, $\bar{m}$, and the proper-time 
regularization cut-off parameter, $\Lambda$.
These parameters are fixed in the mesonic sector by reproducing the empirical
values
 of the pion mass $m_\pi=139$ MeV and the pion decay constant 
\mbox{$f_\pi=93$}~MeV, as explained in the following.

In order to study the vacuum mesonic properties, 
in terms of which the other parameters
will be fixed, we consider the partition function of the model.
It is given by the path integral
\begin{equation}
 Z=\int\!{\cal D}U{\cal D}\psi^\dagger{\cal D}\psi e^{-\int\!d^4z
 \psi^\dagger D(U)\psi }=\int\!{\cal D}Ue^{N_c {\rm Tr}\log[D(U)]}
 \equiv\int\!{\cal D}Ue^{-S_{\rm eff}[U]} 
\end{equation}
where an integration over the quarks  was performed, since they appear
in a quadratic form. The effective action $S_{\rm eff}[U]$ can be 
written, in an unregularized form, as
\begin{equation}
 S_{\rm eff}[\sigma,\vec{\pi}]
 =-N_c{\rm Tr}\log[D(U)]+N_c{\rm Tr}\log[D(U\!=\!1)]+
 \lambda \int\!d^4x(\sigma^2+\vec{\pi}^2-M^2).
\label{eq:seff}\end{equation}
In this expression, explicit scalar and pseudo-scalar meson fields
\begin{equation}
 M U^{\gamma_5}\equiv e^{i\gamma_5\mbox{\boldmath 
 $\scriptstyle\tau$}\cdot \vec{\theta}}
 =\sigma +i\gamma_5 \vec{\pi}\cdot\mbox{\boldmath $\tau$} ,
\end{equation}
are introduced, `Tr' represents a functional trace -- integration 
over space-time and trace over the spin and flavour degrees of freedom -- 
and $\lambda$ is a lagrange multiplier imposing the chiral circle 
condition in the form $\sigma^2+\vec{\pi}^2=M^2$. 
The factor $N_c$ comes from the trace over color and is written
explicitly.
The term ${\rm Tr}\log D(U)$ is the so called one-quark-loop
contribution since $D^{-1}(U)$ is the quark propagator in the 
background of the meson fields. 

The one-quark-loop 
contribution  $N_c {\rm Tr}\log D$ is real in SU(2),
hence a proper-time regularization of (\ref{eq:seff}), $S_{\rm eff}^\Lambda$, 
can be obtained by first 
observing that ${\rm Re}N_c {\rm Tr}\log D=1/2 N_c{\rm Tr}\log(D^\dagger D)$
and then by using (\ref{eq:reg}).
This effective action has a translationally invariant stationary point
at $\sigma_c=M$, $\pi^a_c=0$, identified with the vacuum 
expectation values of the meson fields in the one-quark-loop approximation.
It is a property of the effective action that the inverse 
field propagators are given by its second-order variation with respect
to the auxiliary meson fields at the stationary point.
In momentum space, evaluating the traces in a
plane wave basis~\cite{Jam92} and for the pseudo-scalar field, this
property
\begin{equation}
 \left.{\delta S_{\rm eff}^\Lambda\over \delta\pi(q)\pi(-q)}
 \right|_{q^2=-m_\pi^2}=Z_\pi(q^2=-m_\pi^2)
 \left[q^2+m_\pi^2\right|_{q^2=-m_\pi^2}
\label{eq:pioninvprop}\end{equation}
leads to the identification, with $M^\prime=M+\bar{m}$,
\begin{equation}
 m_\pi^2={\bar{m}\over M^\prime} {\lambda\over Z_\pi(q^2=-m_\pi^2)},
\label{eq:mpi}\end{equation}
which vanishes in the chiral limit. Here,
$\lambda$ can be determined from the so called gap-equation that 
results from the vanishing of the first variation of the action.
Explicitly:
\begin{equation}
  \lambda={4N_c\over (4\pi)^2}\int\limits_0^\infty\!{du\over u^2}
  \phi(u,\Lambda) e^{-u{M^\prime}^2},
\end{equation}
\begin{equation}
  Z_\pi(q^2)={4N_c\over 2(4\pi)^2}\int\limits_0^\infty\!{du\over u}
  \phi(u,\Lambda)\int\limits_0^1\!d\beta e^{-u[{M^\prime}^2
  +\beta(1-\beta)q^2]}.
\end{equation}
The function $\phi(u,\Lambda)$ comes from the proper-time regularization
(\ref{eq:reg}).

Turning to the pion decay constant, it is defined by
the value of the matrix element of the axial current
between the vacuum and a pion state
\begin{equation}
 \langle0|A_{\mu+}(x)|\pi^-,q\rangle=-iq_\mu f_\pi e^{-iq\cdot x}.
\label{eq:piondecay}\end{equation}
The axial current is defined as the first variation 
of the action $\delta S_{\rm eff}\equiv\int\!A_{\mu a}(\partial\alpha_a)$
under the rotation
\begin{equation}
 U^{\gamma_5}\longrightarrow 
 e^{-i\mbox{\boldmath $\scriptstyle\alpha$}\cdot \vec{\tau}\gamma_5/2} 
 U^{\gamma_5} e^{-i\mbox{\boldmath $\scriptstyle\alpha$}\cdot 
 \vec{\tau}\gamma_5/2}.
\end{equation}
From the regularized action we deduce the following form for
the axial current:
\begin{equation}
 A_{\mu a}(x)=M\int\!{d^4q\over (2\pi)^4}d^4y \: e^{iy\cdot q}
 (\partial_\mu\pi^a) \sqrt{Z_\pi(q)}e^{-iq\cdot x}.
\end{equation}
It is written in terms of the physical pion field,
given from (\ref{eq:pioninvprop}) by $\pi^a\rightarrow\pi^a\sqrt{Z_\pi}$. 
The substitution of this expression for the axial current 
in (\ref{eq:piondecay}) and assuming a canonical quantization 
for the pion field yields for the pion decay constant
\begin{equation}
 f_\pi^2=M^2 Z_\pi(q^2=-m_\pi^2).
\label{eq:fpi}\end{equation}

Equations (\ref{eq:mpi},\ref{eq:fpi}) are now used
for a given dynamical mass $M$ to fix $\Lambda$ and $\bar{m}$, in the
one-quark-loop approximation. For reasonable values of $M$, i.e.
$350MeV \le M \le 450MeV$
we obtain $\bar{m}$ close to its phenomenological 
value of $6$ MeV. 
 
In SU(3) there is additionally only the strange quark mass parameter $m_s$, 
since in this approach the strange and non-strange constituent quark masses are
bound to be the same, i.e. $M$.
 The details are more involved and can be found
in~\cite{Blo93}.
 The relations above for the pion are maintained.
In the same line of reasoning as for SU(2), the empirical value 
of the kaon mass can be used to fix $m_s$:
\begin{equation}
m_K^2={\lambda\over Z_K(q^2=-m_K^2)}{\bar{m}+m_s\over M}+(m_s-\bar{m})^2.
\end{equation}
It is found, however, that the value for $m_s$ obtainable
in this way is lower than the 
phenomenological value of $150$ MeV. 
We prefer then to fix $m_s$ to its phenomenological value
by introducing a second parameter $\Lambda_2$ into the function $\phi$
involved in the proper-time regularization (\ref{eq:reg}), in the form
\begin{equation}
 \phi(u,\Lambda_1,\Lambda_2)=
 \theta\left(u-1/\Lambda_1^2\right)\left[
 (u-\Lambda_1)/( \Lambda_2-\Lambda_1)
 +\theta\left(u-1/\Lambda_2^2\right)\right],
\label{eq:phisu3}\end{equation}
which  is then used in the regularization of all quantities in the SU(3) case.

It can be shown~\cite{Chr96,Alk96}
 that, with the effective action considered here, the
Goldstone theorem, the Golberger-Treiman and the Gell-Mann-Oakes relations
are verified. The remaining parameter, the value of the constituent 
quark mass $M$, is 
the only parameter left to be fixed in the baryonic sector. In fact it is
taken from the review \cite{Chr96} with a value of $M=420MeV$. This value
of $M$ has been adjusted to yield in average a good description to many nucleon
observables as e.g. radii, form factors, various charges, etc.

\section{The baryons in the CQSM}

In order to allow for a better understanding of both the way in which
matrix elements of quark operators 
(\ref{ev}) are treated and how the baryons are described in this model,
we start by showing how the soliton mean-field solution can be obtained 
through the nucleon correlation function. 
It will be seen that the saddle-point approximation 
for the correlation function, expressed as a path integral, agrees with
the Hartree picture of the nucleon as a bound state of $N_c$ weakly 
interacting quarks in the meson field $U_c$. Again, all details of the model
are well understood and have been described in Refs.\cite{Dia97,Chr96,Alk96}
with many applications in Ref.~\cite{Chr96}.

However, such mean-field solitonic solutions do not have 
the appropriate quantum numbers to describe a baryon because 
the hedgehog mean-field soliton
$U_{c}(\mbox{\boldmath $x$})$  breaks the translational, rotational and
isorotational symmetries of the action.  These quantum numbers 
are obtained in a {\it projection after variation} way, 
implemented using the path integral~\cite{Dia88}.
\subsection{The mean-field solution: the soliton}
The nucleon correlation function
$ \langle 0|J_{B}(0,T/2)J^\dagger_{B}(0,-T/2)|0\rangle ,$
where the baryonic current
\begin{equation} J_{B}(\mbox{\boldmath $x$},t)=\frac{1}{N_{c}!}
\varepsilon^{\alpha _{1}
\cdots \alpha_{N_c}}\Gamma_{B}^{\{f\}}
\psi _{\alpha _{1}f_{1}}(\mbox{\boldmath $x$},t)\cdots \psi
_{\alpha _{N_{c}}f_{N_{c}}}(\mbox{\boldmath $x$},t) 
\label{curr}\end{equation}
is constructed from $N_c$ quark fields $\psi_{\alpha f }$,
the $\alpha_i$ are color indices, $f_i$ represents both flavour and spin 
indices and $\Gamma_{B}^{\{f\}}$ is a matrix that 
carries the quantum numbers of the baryonic state $B$,
can be represented by the following path integral in euclidean space:
\begin{eqnarray}
\lefteqn{\langle0|J_{B}(0,T/2)J^\dagger_{B}(0,-T/2)|0\rangle 
    } \nonumber \\
 & &\mathrel{\mathop{=}\limits_{T\rightarrow \infty }} 
  {1\over \mathcal{Z}} \int\!{\cal D}\psi^\dagger{\cal D}\psi {\cal D}\, U
  J_{B}(0,T/2)J^\dagger_{B}(0,-T/2) 
  e^{-\int\!d^4z\psi^\dagger D(U)\psi}.
\label{eq:corr}\end{eqnarray}
The large euclidean time separtion $T$ behaviour of this correlation 
function is dominated by the state with lowest energy, 
\begin{equation}
 \langle0|J_{B}(0,T/2)J^\dagger_{B}(0,-T/2)|0\rangle
 \mathrel{\mathop{\sim}\limits_{T\rightarrow \infty }} e^{-M_c T},
\end{equation}
whose determination relies on the evaluation of the right-hand side of
(\ref{eq:corr}) within some set of assumptions.

In (\ref{eq:corr}) the quarks can be integrated exactly 
due to the fact that the action is quadratic in the quark fields.
After that, (\ref{eq:corr}) can be written as
\begin{eqnarray}
  \lefteqn{\langle0|J_{B}(0,T/2)J^\dagger_{B}(0,-T/2)|0\rangle
     } \nonumber \\
& &  \mathrel{\mathop{=}\limits_{T\rightarrow \infty }}
  {1\over \mathcal{Z}} \int\!{\cal D}\, U\,
  \Gamma_B^{\{f\}}\prod\limits_{k}^{N_{c}} 
  \left\langle 0,T/2\left| D^{-1}(U) 
  \right|0,-T/2\right\rangle_{f_k g_k}\Gamma_B^{\{g\}}
   e^{N_c {\rm Tr} \log[D(U)]},
\label{eq:corr2}\end{eqnarray}
where one can recognize the quark propagators coming from the 
contraction of the quark fields in the baryonic currents. 
The spectral representation
for the quark propagator in a static field $U$ is given by
\begin{eqnarray}
 \left\langle \mbox{\boldmath $  x$}^{\prime},x^\prime_4\left| 
   D^{-1}(U) 
 \right|\mbox{\boldmath $  x$},x_4\right\rangle\! &\!=\!&\!
 \theta(x^\prime_4-x_4)\sum_{\varepsilon_n>0} 
 e^{-\varepsilon_n(x^\prime_4-x_4)} 
 \phi_n(\mbox{\boldmath $  x$}^{\prime})
 \phi^\dagger_n(\mbox{\boldmath $  x$}) \nonumber \\
 &\!-\! & \! 
 \theta(x_4-x^\prime_4)\sum_{\varepsilon_n<0} 
 e^{-\varepsilon_n(x^\prime_4-x_4)} 
 \phi_n(\mbox{\boldmath $  x$}^{\prime})
 \phi^\dagger_n(\mbox{\boldmath $  x$}), 
 \label{eq:spectral_rep}
\end{eqnarray}
written in terms of the eigenvalues and eigenfunctions of the 
one-particle Dirac hamiltonian~(\ref{hamiltonian})
\begin{equation}
 h(U)\phi_n(\mbox{\boldmath $x$})
=\varepsilon_n\phi_n(\mbox{\boldmath $x$}).
\label{eq:phi}\end{equation}
Using this spectral representation, it is easy to see that the 
large euclidean time separation $T$ yields for the  product 
of the $N_c$ quark propagators
\begin{equation}
  \prod\limits^{N_{c}} 
  \left\langle 0,T/2\left| D^{-1}(U) 
  \right|0,-T/2\right\rangle
  \mathrel{\mathop{\sim}\limits_{T\rightarrow \infty }}
  e^{-T N_c \varepsilon_{val}(U)},
\end{equation}
where $\varepsilon_{val}$ is the so called valence level, i.e. the bound level
of low positive energy.

The fermion determinant in (\ref{eq:corr2}) 
contains ultraviolet divergencies and must be regularized. Schematically,
\begin{equation}
  \left. N_c {\rm Tr}\log[ D(U)] - N_c {\rm Tr}\log[ D(U\!=\!1)]
  \right|_{reg}
  \mathrel{\mathop{\sim}\limits_{T\rightarrow \infty }}
  -T N_c\!\! \sum_{\varepsilon_n,\varepsilon^{(0)}_n<0}
  \!\!  \left(\varepsilon_n - \varepsilon^{(0)}_n \right)_{reg}
  \equiv -T N_c \varepsilon_{sea}^{reg}(U),
\end{equation}
where the term ${\rm Tr}\log[ D(U\!\!=\!\!1)]$ comes from the normalization
factor $\mathcal{Z}$ in (\ref{eq:corr}) and refers to the vacuum 
configuration $U\!=\!1$. 
The $\varepsilon^{(0)}$
are the eigenstates of the Dirac hamiltonian for this case: 
\begin{equation}
 h(U\!=\!1)\phi^{(0)}_n(\mbox{\boldmath $  x$})
 =\varepsilon^{(0)}_n\phi^{(0)}_n(\mbox{\boldmath $  x$}),
\label{eq:phi0}\end{equation}
with $\phi^{(0)}_n(\mbox{\boldmath $  x$})$ the free
particle wave functions. 

Finally, one obtains
\begin{equation}
  {1\over \mathcal{Z}}\prod\limits^{N_{c}} 
  \left\langle 0,T/2\left| D^{-1}(U) 
  \right|0,-T/2\right\rangle \: e^{N_c {\rm Tr}\log[ D(U)]}
  \mathrel{\mathop{\sim}\limits_{T\rightarrow \infty }}
  e^{-T [N_c \varepsilon_{val}(U)+N_c \varepsilon_{sea}^{reg}(U)]}.
\end{equation}
The saddle point approximation for the integration over $U$ is
now justified by the large $N_c$ limit. 
The corresponding field configuration $U$
with baryon number one is then represented by
\begin{equation} \delta_U\Bigl(N_c\varepsilon_{val}[U]
  +\varepsilon_{sea}[U]\Bigr|_{U=U_c}=0 ,\end{equation} 
which is solved by an iterative self-consistent procedure 
in a finite quasi--discrete basis. Apparently the pion field 
$\vec{\pi}(\mathbf{x})$ in
$U$ is not an independent dynamical field with e.g. independent contributions
to observables. The  $\vec{\pi}(\mathbf{x})$ is basically an 
abbreviation for the
pseudoscalar quark density of the occupied single quark states.
This equation is solved imposing an aditional constraint on the
field $U$, namely an hedgehog shape:  $\theta(\mbox{\boldmath
$x$})=\mbox{\boldmath $\hat{x}$} F(r)  $ with $\mbox{\boldmath $\hat{x}$}=
\mbox{\boldmath $x$}/r$, $r=|\mbox{\boldmath $  x$}|$ and the profile
function of the soliton $F(r)$ satisfying  $F(r)\rightarrow 0$\ as
$r\rightarrow \infty $ and $F(0)=-\pi $. 

\subsection{Quantization} 
The apropriate baryonic quantum numbers are obtained by restricting,
in the path integral, the $U$ field configurations to time 
dependent fluctuations of the field $U_c(\mbox{\boldmath $x$})$ 
along the zero modes. To these modes correspond large amplitude fluctuations
related to the global symmetries of the action: translations and rotations
in space and rotations in flavour space. The two type of rotations are
connected by the hedgehog, though. The small amplitude fluctuations,
corresponding to non-zero modes, correspond to higher terms in the $1/N_c$
expansion and are therefore neglected. 

The large amplitude fluctuations can be treated in the path 
integral formalism. This is achieved by using
\begin{equation} U(\mbox{\boldmath $x$},x_4)
 =A(x_4)\,U_c\Bigl( \mbox{\boldmath $x$} 
 -\mbox{\boldmath $X$}(x_4)\Bigr) A^{\dagger}(x_4)\, , 
\label{eq:zeromodes}\end{equation}
where $A(x_4)$ is a unitary time-dependent  SU(2) or 
SU(3) rotation matrix in flavour space
and $\mbox{\boldmath $X$}(x_4)$ the parameter of a translation.
The integration over $\mbox{\boldmath $X$}$ will provide a
projection into states with definite momentum and the rotation in
flavour space will allow the obtention of states with definite
spin and isospin quantum numbers.

In the context of a rotating stationary field configuration 
$U(\mbox{\boldmath $x$})$,
the operator $D(U)$ is modified according to
\begin{eqnarray} 
  \lefteqn{   D(U)\equiv D\Bigl[ A(x_4)U\bigl(\mbox{\boldmath $x$}
  \!-\mbox{\boldmath $X$}(x_4)\bigr)A^\dagger (x_4)\Bigr]  }\nonumber\\
& & \mbox{\hspace{1cm}}=A\:e^{-i\mbox{\boldmath $\scriptstyle P$}\cdot 
            \mbox{\boldmath $\scriptstyle X$}}
    \Bigl(D(U_c) +A^\dagger\dot{A}
     -i\mbox{\boldmath $P$}\cdot \dot{\mbox{\boldmath $X$}} 
    + \gamma_4 A^\dagger\delta m\, A \Bigr)
     e^{i\mbox{\boldmath $\scriptstyle P$}\cdot 
             \mbox{\boldmath $\scriptstyle X$}} A^\dagger  ,
\label{d}    \end{eqnarray}
in which $\delta m=m-\bar{m}\mathbf{1}$ is absent in SU(2).

Considering the ansatz (\ref{d}) in the expression
(\ref{eq:corr2}) results in 
\begin{eqnarray}
  \lefteqn{\langle0|J_{B}(0,T/2)J^\dagger_{B}(0,-T/2)|0\rangle
  \mathrel{\mathop{=}\limits_{T\rightarrow \infty }} 
  {1\over \mathcal{Z}} \int\!d^3 X\int\!\!{\cal D}\, A\,\: \Gamma_B^{\{f\}}  } 
     \nonumber \\
& & \times \prod\limits_{k}^{N_{c}} 
  \langle -\mbox{\boldmath $X$},T/2|
  A(T/2)\:[D(U_c) +A^\dagger\dot{A}]^{-1}\: A^\dagger(-T/2)\;
  |-\mbox{\boldmath $X$},-T/2\rangle_{f_k g_k} \nonumber \\
& &\times \:  \Gamma_B^{\{g\}}\:
   e^{N_c {\rm Tr} \log[D(U_c)+A^\dagger\dot{A}]}.
\label{eq:corr3}\end{eqnarray}
The significant change is 
the substitution of the integration over $U$  by an integration 
over the matrices $A$ specifying the orientation of the soliton 
in flavour space.

The term $A^\dagger\dot{A}$ allows for the introduction of an hermitian
angular velocity matrix~\cite{Blo93}
\begin{equation}
 \Omega=-iA^\dagger\dot{A}\equiv\sum_a\Omega_a t_a,
\end{equation}
standing $t_a$ for $\tau_a/2$ in SU(2) and $\lambda_a/2$
in SU(3). The next step consists in assuming adiabatic rotation and performing
an expansion in $\Omega$ treating it
 as small and neglecting its derivatives.
This is justified because from the delta--nucleon mass splitting one can
estimate that
 $\Omega$ is $ O(1/N_c)$. 
 Using
\begin{equation}
 (D(U_c)+i\Omega)^{-1} = D^{-1}(U_c) - D^{-1}(U_c)i\Omega D^{-1}(U_c) + \cdots
\label{eq:omegaexp}\end{equation}
the product of the $N_c$ propagators and fermion determinant becomes
\begin{eqnarray}
\lefteqn{  \mbox{\hspace{-1cm}} {1\over Z}\prod\limits^{N_{c}} 
  \left\langle -\mbox{\boldmath $X$},T/2\left| D^{-1}(U) 
  \right|-\!\mbox{\boldmath $X$},-T/2\right\rangle_{f_i g_i} 
  \: e^{N_c {\rm Tr}\log[ D(U)]}
   }  \nonumber\\
   &   &  \mathrel{\mathop{\sim}\limits_{T\rightarrow \infty }}
   A_{f_i f^\prime_i}
   \phi^{\rm val}_{f_i^\prime}(\mbox{\boldmath $X$})
   {\phi^{\rm val}}^\dagger_{g_i^\prime}(\mbox{\boldmath $X$})
   A_{g^\prime_i g_i}
   e^{-T M_c-\int\!dx_4 L_{\rm rot} }.
\label{eq:corr6}\end{eqnarray}
The rotational lagrangian is given by
\begin{equation}
 L_{\rm rot}={1\over2}\sum_{a=1}^3 I_1\Omega_a^2 +
 {1\over2}\sum_{a=3}^7 I_2\Omega_a^2  - {N_c\over2\sqrt{3}}\Omega_8
\label{eq:lagrot}\end{equation}
where the two last terms concern SU(3) only.
This SU(3) result is obtained from (\ref{eq:corr6})
through the projection onto strange and non-strange subspaces
(\ref{eq:dsu3}). The moments of inertia $I_1$, $I_2$ 
are $O(1/N_c)$ and are regularized quantities.

The expression (\ref{eq:corr6}) shows that, in the large 
$N_c$ limit, the integration over the orientation matrices 
of the soliton $A$ is dominated by those trajectories which are close 
to the ones of the quantum spherical rotator with collective hamiltonian
\begin{equation}
 H_{\rm coll}={1\over2I_1}\sum_{a=1}^3 J_a^2 +
 {1\over2I_2}\sum_{a=3}^7 J_a^2 
\label{eq:hamilcoll}\end{equation}
with the $J_a$, $a=1,2,3$ identified as spin operators
and playing, in this context, the part of right rotation 
generators~\cite{Blo93}.
The quantization rules that follow are
\begin{equation}
i\Omega_a=\left\{\begin{array}{cc} J_a/I_1 & ,a=1,2,3 \\
                                   J_a/I_2 & ,a=4,\ldots,7 \end{array}
\right. \end{equation} 
and $J_8=-N_c/(2\sqrt{3})$, which can be read from the linear 
term in $\Omega_8$ in (\ref{eq:lagrot}).
This constraint in $J_8$ can be cast in terms of the `right' hypercharge $Y_R$,
which, in analogy with the hypercharge, can be defined as $Y_R=2J_8/\sqrt{3}=
-N_c/3$. Constraining it to be $-1$ also constrains the SU(3) representantions
to the octet and decuplet with spins $1/2$ and $3/2$ respectively.

Concerning the wave functions, it is possible to show~\cite{Chr96} that
the contraction of a matrix $\Gamma_B\equiv\Gamma_{JJ_3TT_3}$, carrying the
quantum numbers of the baryon, with 
the product of $N_c$ rotation matrices $A$ and valence wave functions
$\phi^{val}(\mbox{\boldmath $X$})$, which have grand spin $0$, result formally in
\begin{eqnarray}
\lefteqn{  \sum_{\{f_k,f_k^\prime\}}\Gamma_{JJ_3TT_3}^{\{f_k\}}
  A_{f_1f_1^\prime}(T/2)\cdots A_{f_{N_c}f_{N_c}^\prime}(T/2)
  \int\!d^3X \phi^{val}_{f_1}(\mbox{\boldmath $X$})\ldots
  \phi^{val}_{f_{N_c}}(\mbox{\boldmath $X$})  }\nonumber\\
& = & \sum_{T_3^\prime} \sum_{\{f_k^\prime\}}\Gamma_{JJ_3TT_3}^{\{f_k\}}
  D^{(T)}_{T_3T_3^\prime}\bigl[A(T/2)\bigr]
  \int\!d^3X \phi^{val}_{f_1}(\mbox{\boldmath $X$})\ldots
  \phi^{val}_{f_{N_c}}(\mbox{\boldmath $X$}) \nonumber \\
& \longrightarrow  & (-1)^{J+J_3}D^{(T)}_{T_3,-J_3}\bigl[A(T/2)\bigr].
\end{eqnarray}
Apart from a normalization factor, this is the wave function of 
the collective state written in terms of the Wigner $D$ functions, 
given in SU(2) by
\begin{equation}
\psi_B(A)\equiv\psi _{TT_{3},JJ_{3}}(A)=\sqrt{2T+1}
(-)^{T+J_{3}}D_{-T_{3}J_{3}}^{(T=J)}(A).
\end{equation}
This function is an eigenfunction of the hamiltonian (\ref{eq:hamilcoll}),
as expected. In SU(3), analogously, the wave function is given by
\begin{equation}  \psi _{(YTT_{3}),(Y^{\, \prime }=-1\,JJ_{3})}^{(n)}(A)=
    \sqrt{\dim (n)}(-)^{Y^{\prime }/2+J_{3}}D_{(YTT_{3}),
    (Y^{\prime }=-1,J,-J_{3})}^{(n)\ast }(A).               \end{equation}

The path integration over $A$ in (\ref{eq:corr3}) can now be 
carried out using
\begin{equation}
 \int\!{\cal D}A\;\psi^*_B(A)\psi_B(A)e^{-I/2\int dx_4\Omega^2}\equiv
 \langle B|e^{-H_{\rm coll}}T|B\rangle =
 \int\!dA\, \psi^*_B(A)\psi_B(A) e^{-TJ(J+1)/(2I)}.
\end{equation}
From such formalism, the Nucleon-$\Delta$ mass splitting is 
easely computed and found to reproduce well the experimental value. Actually,
although the SU(3) formalism is obtained by assuming the famous embedding
\cite{Witten}, the results of SU(2) and SU(3) are by no means identical since
the flavour rotation is performed in different flavour spaces. 

\subsection{Baryonic matrix elements} 
The baryon expectaction values of quark currents, $\bar{\psi}O\psi $, 
being  $O$ some matrix with spin and isospin indices, 
can be expressed as a functional integral~\cite{Dia88} through
\begin{eqnarray}
\lefteqn{  \left\langle B^\prime,\mbox{\boldmath $p$}^\prime\left| 
  \bar{\psi}O\psi 
  \,\right| B,\mbox{\boldmath $p$}\right\rangle 
  =\lim_{T\rightarrow \infty } 
  \frac{1}{\cal Z} \int\! d^3 x\,d^3 x^{\prime }\:
  e^{-i(\mbox{\boldmath $\scriptstyle p$}^{\prime }\cdot
  \mbox{\boldmath $\scriptstyle x$}^{\prime}-
  \mbox{\boldmath $\scriptstyle p$}\cdot
  \mbox{\boldmath $\scriptstyle x$})}      }\nonumber \\
& & \times \int\! {\cal D}U{\cal D}\bar{\psi}{\cal D}\psi \,\,J_{B^{\prime }}
  \left( \mbox{\boldmath $x$}^{\prime },T/2\right) \,
  \bar{\psi} O\psi \,J_{B}^\dagger\left( \mbox{\boldmath $x$},-T/2\right) 
   e^{-\int\! d^{4}z\,\,\psi^\dagger D(U)\psi } ,
\label{ev}  \end{eqnarray}
in which the baryonic state is again created from the vacuum by the current 
$J_{B}^\dagger$ given by (\ref{curr}).
When the quarks are integrated out in (\ref{ev}) the result
is given as a sum of two parts: a valence part,
\begin{eqnarray}
\lefteqn{ \left\langle B^\prime,\mbox{\boldmath $p$}^\prime
  \left| \bar{\psi}O\psi \,
  \right| B,\mbox{\boldmath $p$}\right\rangle _{val}  
  =\lim_{T\rightarrow \infty } 
  N_c\frac{1}{\cal Z}   \int\!\! d^3x\,d^3x^{\prime }\,
  e^{-i(\mbox{\boldmath $\scriptstyle p$}^{\prime }\cdot 
        \mbox{\boldmath $\scriptstyle x$}^{\prime }-
        \mbox{\boldmath $\scriptstyle p$}\cdot 
        \mbox{\boldmath $\scriptstyle x$} )}
  \:\Gamma_{B^{\prime }}^{\{f\}\,}  \Gamma_{B}^{\{g\}} 
   } \mbox{\hspace{2cm}}\nonumber\\   
& &\times  
  \int\! {\cal D}A\: e^{N_c\, {\rm Tr}\, \log \, D(U)}
  \:\prod\limits_{k}^{N_{c}} 
  \left\langle \mbox{\boldmath $x$}^{\prime },T/2\left| D^{-1}(U) 
  \right|\mbox{\boldmath $x$},-T/2
  \right\rangle_{f_{k}g_{k}} \nonumber\\
&&\times  \left\langle 
  \mbox{\boldmath $x$}^{\prime} T/2\left| D^{-1}(U)
  \right|0,0\right\rangle_{f_{1}d}^{{}} O_{dd^\prime}
  \left\langle 0,0\left|D^{-1}(U) \right|\mbox{\boldmath $x$}
  ,-T/2\right\rangle_{d^{\prime }g_{1}}  
  \, ,  \label{eq:val}
\end{eqnarray}
and a Dirac sea part,
\begin{eqnarray}
\lefteqn{ \left\langle B^\prime,\mbox{\boldmath $p$}^\prime
  \left| \bar{\psi}O\psi \,
  \right| B,\mbox{\boldmath $p$}\right\rangle _{sea} 
  =\lim_{T\rightarrow \infty } 
  N_c \frac{1}{\cal Z}   \int\!\! d^3x\,d^3x^{\prime }\,
  e^{-i(\mbox{\boldmath $\scriptstyle p$}^{\prime }\cdot 
        \mbox{\boldmath $\scriptstyle x$}^{\prime }-
        \mbox{\boldmath $\scriptstyle p$}\cdot 
        \mbox{\boldmath $\scriptstyle x$} )}
  \:\Gamma_{B^{\prime }}^{\{f\}\,}  \Gamma_{B}^{\{g\}} }
  \mbox{\hspace{2cm}}\nonumber\\   
&&\times  
  \int\! {\cal D}A\:\prod\limits_{k}^{N_{c}} 
  \left\langle \mbox{\boldmath $x$}^{\prime },T/2\left| D^{-1}(U) 
  \right|\mbox{\boldmath $x$},-T/2\right\rangle_{f_{k}g_{k}} \nonumber\\
&&\times {\rm Tr}\left\{ O_{dd^\prime}
  \left\langle 0, 0  \left|- D^{-1}(U) 
  \right|0,0 \right\rangle_{dd^\prime} \right\} 
  \: e^{N_c\, {\rm Tr}\, \log \, D(U)} .  \label{eq:sea}
\end{eqnarray}

Proceeding in the same way as in the preceding section,
the valence contribution after some simple manipulations
acquires the form~\cite{Chr95,Kim96a} in SU(2)
\begin{equation}
  \left\langle B^\prime,\mbox{\boldmath $p$}^\prime
  \left| \bar{\psi}O\psi \,
  \right| B,\mbox{\boldmath $p$}\right\rangle _{val}  
  = 
  N_c   \int\!\! d^3x\,
  e^{-i(\mbox{\boldmath $\scriptstyle p$}^{\prime }-
        \mbox{\boldmath $\scriptstyle p$})\cdot 
        \mbox{\boldmath $\scriptstyle x$} }
  \int\! dA \psi_{B^\prime}^\ast(A) \bigl[{\cal V}^{(\Omega^0)}
  (\mbox{\boldmath $x$}) + {\cal V}^{(\Omega^1)}
  (\mbox{\boldmath $x$}) \bigr] \psi_{B}(A)
\label{eq:densis}\end{equation}
with a leading term, independent of the angular velocity ($\Omega^0$),
\begin{equation}
 {\cal V}^{(\Omega^0)} (\mbox{\boldmath $ x$})=
  \phi^\dagger_{\rm val}(\mbox{\boldmath $ x$})
  A^\dagger O A \: \phi_{\rm val}(\mbox{\boldmath $ x$}),
\label{eq:densi0}\end{equation}
and a term proportional to the angular velocity ($\Omega^1$), hence
 $O(1/N_c)$, 
\begin{eqnarray}
\lefteqn{  {\cal V}^{(\Omega^1)} (\mbox{\boldmath $x$})=
 {1\over 2I_1}\sum\limits_{\varepsilon_n\neq\varepsilon_{\rm val}} 
 {1\over \varepsilon_{\rm val}-\varepsilon_n}  }\nonumber\\
 &  \times & \Bigl\{ \theta(\varepsilon_n) \left[ J_a\phi^\dagger_n
 (\mbox{\boldmath $x$})
  A^\dagger O A \, \phi_{\rm val}(\mbox{\boldmath $x$})
 \langle{\rm val}|\tau^a|n\rangle
 +\phi^\dagger_{\rm val}(\mbox{\boldmath $x$})
  A^\dagger O A\,  \phi_n(\mbox{\boldmath $x$})
 \langle n|\tau^a|{\rm val}\rangle J_a \right] \nonumber\\
 & &+ \theta(-\varepsilon_n)  
 \left[ J_a \phi^\dagger_{\rm val}(\mbox{\boldmath $x$})
  A^\dagger O A \, \phi_n(\mbox{\boldmath $x$})
 \langle n|\tau^a|{\rm val}\rangle
 +\phi^\dagger_n(\mbox{\boldmath $x$})
  A^\dagger O A \, \phi_{\rm val}(\mbox{\boldmath $x$})
 \langle{\rm val}|\tau^a|n\rangle J_a \right] \Bigr\} .
\label{eq:densi}\end{eqnarray}
For the SU(3) flavour case the structure is the same 
except for the existence of more terms resulting from the
projection (\ref{eq:dsu3}).

The last important aspect  is
the ordering of the collective operators, which accounts for the
different orderings of $J_a$ and $A^\dagger O A$
present in (\ref{eq:densi}). In the non-singlet case, writing the
operator $O$ in the form $O=O^a\tau^a$ gives
\begin{equation} 
 A^\dagger \tau^a A = D^{(1)}_{ab}(A)\tau^b ,
\end{equation}
with the definition $D^{(1)}_{ab}(A)=(1/2)\,{\rm tr}\,
(A^\dagger\tau^aA\tau^b),$
which does not comute with the spin operators $J_a$. Therefore, the proper
time ordering of the collective operators must be taken into account 
in arriving at (\ref{eq:densi}), both in SU(2) and SU(3)~\cite{Chr96}.
\section{Electroproduction of the $\Delta$(1232)} 
Now we turn to the problem of the $\Delta$--electroproduction. 
One should note
here, on the basis of the preceeding sections,
that in the present formalism the $\Delta$ is a bound state which
corresponds to a soliton rotating in flavour space. Hence it is as stable as
the nucleon and does not decay in nucleon and pion without strong
modification of the model. 

The electromagnetic current, obtained by minimally coupling
the photon field with the quarks at the level of the
lagrangian (\ref{lagrangian}), is conserved. 
However, that may not be the case in calculations based 
on the $1/N_c$ expansion, since a truncation is always involved and,
in the end, $N_c$ is taken to be $N_c=3$. 
In the present case, we have further to add that
not all the possible $1/N_c$ contributions are included:
the ansatz  (\ref{eq:zeromodes}) does not contain
modes orthogonal to the zero modes. 
We follow the  assumption 
that the contribution of such modes, of higher order in $1/N_c$ 
is small, as is in the case for 
other baryonic observables in the framework used here~\cite{Chr96}.
 
The reference frame in which we chose to 
compute the nucleon to $\Delta$ transition amplitude is the rest frame of the
$\Delta$. The kinematics is  then specified by  the nucleon
$(E_{N},-\mbox{\boldmath $q$})$ and photon  
$(\omega ,\mbox{\boldmath $q$})$ four momenta. In terms
of the photon virtuality, $Q^{2}=-q^{2}$, one can further write 
\begin{equation} 
 |\mbox{\boldmath $q$}|^2=\left( \frac{m_\Delta^2+m_N^2+Q^2}{2\,m_\Delta }    
 \right)^2-m_N^2\,          
\end{equation} 
and 
\begin{equation}  
 \omega=\frac{m_\Delta^2-m_N^2-Q^2}{2\,m_\Delta}.       
\end{equation}

The helicity transversal ($A_{\lambda }$) and scalar
($S_{1/2}$) amplitudes are defined by
\begin{equation}
      A_{\lambda}(q^2)=
  -\frac{e}{\sqrt{2\omega}}
  \left\langle \Delta{\scriptstyle \left(\frac{3}{2},\lambda\right)}
  \, \right| \int d^3x
  \,\bar{\psi}{\cal Q}\mbox{\boldmath $\gamma$}\psi \cdot \xi_{+1}
  e^{i\mbox{\boldmath $\scriptstyle q$}
  \cdot\mbox{\boldmath $\scriptstyle x$}} \left| 
  N{\scriptstyle \left(
  \frac{1}{2},\lambda -1\right)}\right\rangle 
\label{amplitudea}  \end{equation}
and
\begin{equation}
  S_{1/ 2}(q^2) = -\frac{e}{\sqrt{2\omega}}\frac{1}{\sqrt{2}}
  \left\langle \Delta {\scriptstyle \left(\frac{3}{2},\frac{1}{2}\right)}
   \right|\int d^3x \,\bar{\psi}{\cal Q}\gamma^0\psi  
  e^{i\mbox{\boldmath $\scriptstyle q$}
  \cdot\mbox{\boldmath $\scriptstyle x$}} \left|
  N {\scriptstyle \left(\frac{1}{2},\frac{1}{2}\right)} \right\rangle \, ,
\label{amplitudes}  \end{equation}
where $\xi_{+1}=-1/\sqrt{2}(1,+i,0)$,
$\lambda =1/2,3/2$, the replacement of $1/\sqrt{2\omega}$ by 
$1/\sqrt{2\omega(q^2=0)}$ has been made and 
${\cal Q}$ is the charge matrix.

These amplitudes can be  multipole expanded, leading to 
the following multipole quantities relevant for $\Delta$ electroproduction:
\begin{eqnarray}
{\cal M}^{M1} &=&i\sqrt{6\pi}\int d^{3}x\,\left\langle \Delta
            {\scriptstyle \left({3\over 2},{1\over 2}\right)}
            \right|\left\{Y^1\otimes J^{(1)}\right\}_{11} 
            j_1(|\mbox{\boldmath $q$}|r) \,\left|N{\scriptstyle 
            \left({1\over 2},-{1\over 2}\right)}\right\rangle \, , 
\label{eq:mm1}\\
{\cal M}^{E2} &=&\frac{\sqrt{5\pi }}{3}\frac{\omega }{|\mbox{\boldmath $q$}|}
      \int\! d^{3}x\,\left\langle \Delta{\scriptstyle 
      \left({3\over 2},{1\over 2}\right)} \right| \,
      \rho(\mbox{\boldmath $x$}) Y_{21}(\hat{x})\frac{\partial }{\partial r} 
      \left(rj_{2}(|\mbox{\boldmath $q$}|r)\right)\, \left|
      N{\scriptstyle \left({1\over 2},-{1\over 2}\right)}\right\rangle  
\nonumber \\
    &\mbox{}- &i\frac{\sqrt{5\pi }}{3} |\mbox{\boldmath $q$}| 
    \int\!d^{3}x\,\left\langle \Delta{\scriptstyle 
    \left({3\over 2},{1\over 2}\right)} \right|
    \,\hat{x}\cdot \mbox{\boldmath $J$}\: Y_{21}(\hat{x}) 
    j_{2}(|\mbox{\boldmath $q$}|r)\left| 
    N{\scriptstyle \left({1\over 2},-{1\over 2}\right)}\right\rangle\, ,
 \label{eq:me2} \\
{\cal M}^{C2} &=&-\sqrt{20\pi }\int\!d^{3}x\,\left\langle \Delta
       {\scriptstyle \left({3\over 2},{1\over 2}\right)} 
       \right|\,\rho(\mbox{\boldmath $x$}) j_{2}(|\mbox{\boldmath $q$}|r)\, 
       Y_{20}(\hat{x})\left|
      N{\scriptstyle \left({1\over 2},{1\over 2}\right)} \right\rangle \, . 
\label{eq:mc2}
\end{eqnarray}
These quantities are now in a form suitable to be calculated in the model
applying the formalism described in the preceding section.

The final expressions for the quadrupole electric and scalar multipole
quantities are:
\begin{equation} 
  { {\cal M}_{\rm SU(3)}^{E2} \over 
        \left\langle \Delta \right| D_{Q3}^{(8)}\,\left|N\right\rangle }
   =   2 { {\cal M}_{\rm SU(2)}^{E2} \over 
      \left\langle \Delta \right| D_{00}^{(1)}\,\left|N\right\rangle }
   = -\frac{3}{8\sqrt{2} I_1} {\omega\over |\mbox{\boldmath $q$}| } 
        \int\! d^3x \,\,{\partial\over \partial r }\Bigl(r j_2
      (|\mbox{\boldmath $q$}| r)\Bigr)\,\, {\cal G}^{(\Omega^1)}
   (\mbox{\boldmath $x$})   ,
\end{equation}
and
\begin{equation}  
  { {\cal M}_{\rm SU(3)}^{C2} \over
     \left\langle \Delta \right| D_{Q3}^{(8)}\,\left|N\right\rangle }
  =2 { {\cal M}_{\rm SU(2)}^{C2}  \over
     \left\langle \Delta \right| D_{00}^{(1)}\,\left|N\right\rangle }
  = -\frac{3}{4 I_1} \int\!d^3x\,\,j_2(|\mbox{\boldmath $q$}| r) 
    \,\,{\cal G}^{(\Omega^1)}(\mbox{\boldmath $x$}) .
\end{equation}
The notation $\langle\Delta|D_{ab}| N\rangle$ 
applies to  the integration over the 
collective wave functions
\begin{equation}  
 \left\langle \Delta \right| D_{ab}^{(n)}
 \,\left| N\right\rangle =\int\! dA \,\psi _{\Delta }^{\ast }
 (A)D_{ab}^{(n)}(A)\psi_N (A)\, ,  
\end{equation}
with $\Delta$ and $N$ as shorthand for spin and isospin quantum numbers 
of the baryonic state and  
$D_{Qa}^{(8)}=\lambda_a (D_{3a}^{(8)}+{1\over\sqrt{3}}D_{8a}^{(8)})/2$,
which comes from the rotation of the charge matrix $Q$ in SU(3) flavour
space, $A^\dagger{\cal Q}A=D_{Qa}^{(8)}\lambda_a$.

The density ${\cal G}^{(\Omega^1)}(\mbox{\boldmath $x$})$ is given by
\begin{eqnarray}
{\cal G}^{(\Omega^1)}(\mbox{\boldmath $x$})&=& 
    \sum_{n\neq val}\frac{1}{\varepsilon _{n}-\varepsilon _{val}} 
    \phi^\dagger_n(\mbox{\boldmath $x$}) 
    \{ Y_2\otimes\tau_1\}_{1a} 
    \phi_{\rm val}(\mbox{\boldmath $x$}) 
    \left\langle n| \tau_a|val\right\rangle  \nonumber \\
&+& \sum_{m,n}{\cal R}_{A}(\varepsilon _{m},\varepsilon _{n}) 
    \phi^\dagger_m(\mbox{\boldmath $x$}) 
    \{Y_2\otimes \tau_1\}_{1a} \phi_n(\mbox{\boldmath $x$}) 
    \left\langle m|\tau_a|n\right\rangle 
\end{eqnarray}
with $\left\langle m|\tau_a|n\right\rangle=\int d^3x \phi^\dagger_m
\tau_a\phi_n$ in terms of the hamiltonian eigenfunctions
(\ref{eq:phi}). The regularization function 
${\cal R}_{A}(\varepsilon _{m},\varepsilon _{n})$ is given in
Appendix A.
The comparison with (\ref{eq:densis}) shows that
the leading ($\Omega^0$) term vanishes both in SU(2) and SU(3), that is,
the quadropole amplitudes are $O(1/N_c)$.
Within the present embedding treatment of SU(3),
the only difference between SU(3) and SU(2) comes from the
collective parts because the contribution containing the 
unpolarized strange quark one-particle
states $\phi^{(0)}(\mbox{\boldmath $x$})$ 
vanishes in SU(3), which is not the case for 
${\cal M}_{\rm SU(3)}^{M1}$ below.

As for the the magnetic dipole quantities, we obtain
\begin{eqnarray}
{\cal M}_{\rm SU(3)}^{M1} &=& 
    -\frac{3\sqrt{3}}{2}\left\langle \Delta \right| D_{Q3}^{(8)}\, \left|
    N\right\rangle{\cal F}^{(\Omega^0)}_1(|\mbox{\boldmath $q$}|)
    -\frac{1}{2\sqrt{2}\,I_{1}}\left\langle \Delta \right| D_{Q3}^{(8)}\, \left|
    N\right\rangle {\cal F}^{(\Omega^1)}_2(|\mbox{\boldmath $q$}|) \nonumber  \\
&+& \frac{\sqrt{3}}{4\,I_{2}}\left\langle \Delta \right|
    d_{Qab}D_{Qa}^{(8)}J_{b}\delta_{ab} \, \left| N\right\rangle 
    {\cal F}^{(\Omega^1)}_3(|\mbox{\boldmath $q$}|) 
\end{eqnarray}
and
\begin{equation}  {\cal M}_{\rm SU(2)}^{M1} = 
    -\frac{\sqrt{3}}{2}\left\langle \Delta \right| 
    D_{00}^{(1)}\,\left|N\right\rangle 
     {\cal F}^{(\Omega^0)}_1(|\mbox{\boldmath $q$}|)
     + \frac{\sqrt{2}}{8\,I_{1}}\left\langle \Delta \right| D_{00}^{(1)}
    \,\left|N\right\rangle{\cal F}^{(\Omega^1)}_2(|\mbox{\boldmath $q$}|) \,  
\end{equation}
with
\begin{eqnarray}
   {\cal F}^{(\Omega^0)}_1 (|\mbox{\boldmath $q$}|)&=&
   \int\!d^3x\,j_{1}(|\mbox{\boldmath $q$}| r)
   \Bigl[ \phi^\dagger_{\rm val}(\mbox{\boldmath $x$})
   \gamma _{5}(\hat{\mbox{\boldmath $x$}}
   \times\mbox{\boldmath $\sigma$})\cdot\mbox{\boldmath $\tau$}
   \phi_{\rm val}(\mbox{\boldmath $x$})     \nonumber \\
&+&\sum_{n}{\cal R}_{1}(\varepsilon _{n}  )
   \phi^\dagger_n(\mbox{\boldmath $x$})
   \gamma _{5}(\hat{\mbox{\boldmath $x$}}
   \times\mbox{\boldmath $\sigma$})\cdot\mbox{\boldmath $\tau$}
   \phi_n(\mbox{\boldmath $x$}) \Bigr], 
\end{eqnarray}
\begin{eqnarray}
    {\cal F}^{(\Omega^1)}_2(|\mbox{\boldmath $q$}|)&=& -{1\over2}
    \int\!d^3x\,j_{1}(|\mbox{\boldmath $q$}| r)\Bigl[\sum_{n\neq val}
    \frac{sgn(\varepsilon _{n})}{\varepsilon _{n}-\varepsilon _{val}} 
    \phi^\dagger_n(\mbox{\boldmath $x$})
    i\gamma _{5}\Bigl(
    (\hat{\mbox{\boldmath $x$}}\times\mbox{\boldmath $\sigma$})
    \times\mbox{\boldmath $\tau$}\Bigr)_a
    \phi_{\rm val}(\mbox{\boldmath $x$})
    \left\langle n|\tau_a|val\right\rangle \nonumber  \\
&+& {1\over 2} \sum_{m,n}{\cal R}_{B}(\varepsilon _{m},
    \varepsilon _{n}) 
    \phi^\dagger_m(\mbox{\boldmath $x$})
    i\gamma _{5}\Bigl(
    (\hat{\mbox{\boldmath $x$}}\times\mbox{\boldmath $\sigma$})
    \times\mbox{\boldmath $\tau$}\Bigr)_a
    \phi_n(\mbox{\boldmath $x$})
    \left\langle m|\tau_a|n\right\rangle  \Bigr] 
\end{eqnarray}
and
\begin{eqnarray}
   {\cal F}^{(\Omega^1)}_3(|\mbox{\boldmath $q$}|)&=&
   \int\!d^3x\,j_{1}(|\mbox{\boldmath $q$}| r)\Bigl[\sum_{n^0}
   \frac{1}{\varepsilon^0_n-\varepsilon_{val}} 
   \phi^\dagger_{\rm val}(\mbox{\boldmath $x$})
   \gamma _{5}(\hat{\mbox{\boldmath $x$}}
   \times\mbox{\boldmath $\sigma$})\cdot\mbox{\boldmath $\tau$}
   \phi^{(0)}_n(\mbox{\boldmath $x$})
   \left\langle n^0|val\right\rangle \nonumber \\
&+&\sum_{m,n^0} {\cal R}_{\cal M}(\varepsilon_m,\varepsilon^0_n)
   \phi^\dagger_m(\mbox{\boldmath $x$})
   \gamma _{5}(\hat{\mbox{\boldmath $x$}}\times\mbox{\boldmath $\sigma$})
   \cdot\mbox{\boldmath $\tau$}  
   \phi^{(0)}_n(\mbox{\boldmath $x$})
   \langle n^0|m\rangle  \Bigr]  ,
\end{eqnarray}
with $\left\langle n^0|m\right\rangle=\int d^3x
{\phi^{(0)}}^\dagger_n\phi_m$
and $\phi^{(0)}_n(\mbox{\boldmath $x$})$ given by (\ref{eq:phi0}).
The regularization functions 
${\cal R}_{1}(\varepsilon _{n}  )$,
${\cal R}_{B}(\varepsilon _{m},\varepsilon _{n})$ 
are given in  Appendix A,
and ${\cal R}_{\cal M}$ is given by
\begin{equation}  
     {\cal R}_{\cal M}(\varepsilon_m,\varepsilon_n)
     = \frac{\frac{1}{2}[sgn(\varepsilon_m)-
     sgn(\varepsilon_n)]}{\varepsilon_n-\varepsilon _m}     \, . 
\end{equation}
\section{Results and discussion}
The ratios E2/M1 and C2/M1 are calculated, exactly in the way described
in the previous section, for a 
constituent mass $M$ of 420 MeV, which, after reproducing masses and decay
constants in the mesonic sector, is the only free parameter left to be fixed in
the baryonic sector. For $M$ we chose the canonical value of 420 MeV for which
the chiral quark-soliton model is known to reproduce well \cite{Chr96}
nucleon  observables, like non-transitional form factors, both in 
SU(2)~\cite{Chr95} and SU(3)~\cite{Kim96a}.  The $\Delta$ is also well
described within exactly the same framework. In particular, the
nucleon-$\Delta$ mass splitting is well reproduced~\cite{Chr96}, supporting
the above procedure adopted in calculating observables.

In (\ref{d}), the $\delta m$  term  is
often treated perturbatively in this model. Such a perturbative
expansion in $\delta m$ was not performed in the present paper since it
was found in many calculations in the CQSM  that
 the linear corrections $O(\delta m)$ are in general small~\cite{Chr96}, as
e.g. in the case of magnetic moments~\cite{Kim96}. 
Only for very
sensitive quantities directly related to the strange content of the nucleon
has the inclusion of the term $\delta m$ an effect larger than about 10
percent.

In this calculation, because we aim to study the electroproduction at low
$Q^2$, no correction for relativistic recoil
effects was taken into account explicitly. Such terms are of higher order
in $1/N_c$ and are expected to become important around and above  $1$ GeV$^2$.
 In our approach, the interval for the
photon virtuality is determined by the fact that the
momentum transfer is $O(N_c^0)$, hence parametrically 
limited by the nucleon mass.

\begin{figure}
\hspace{2cm}
\epsfxsize=10.5cm
\epsffile{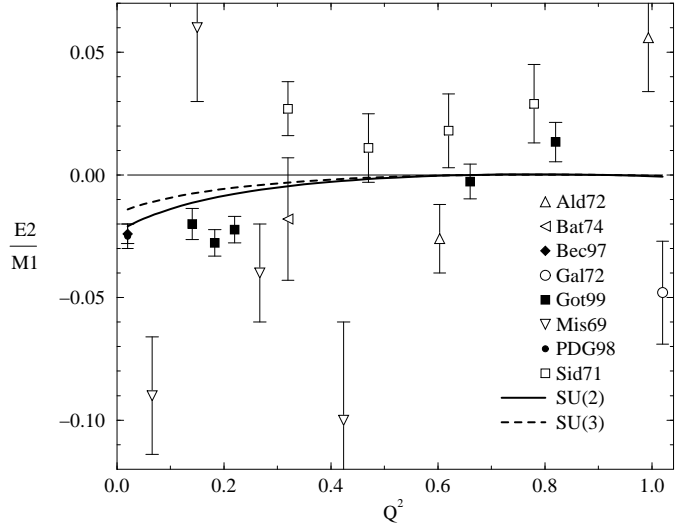}
\caption[]{\protect\small The ratio E2/M1 calculated in the CQSM, 
   in flavor SU(2) and SU(3), 
   for a constituent quark mass of 420 MeV  represented 
   as a function of $Q^2$ in GeV$^2$. 
   Older experimental data (open symbols) 
   is taken from references~\cite{old_data1,old_data2}, more recent data
   (filled symbols) is from~\cite{Got99} and   
   at the photon point from~\cite{photon_point}.  }    
\protect\label{fig:e2m1}
\end{figure}

\begin{figure}
\hspace{2cm}
\epsfxsize=10.5cm
\epsffile{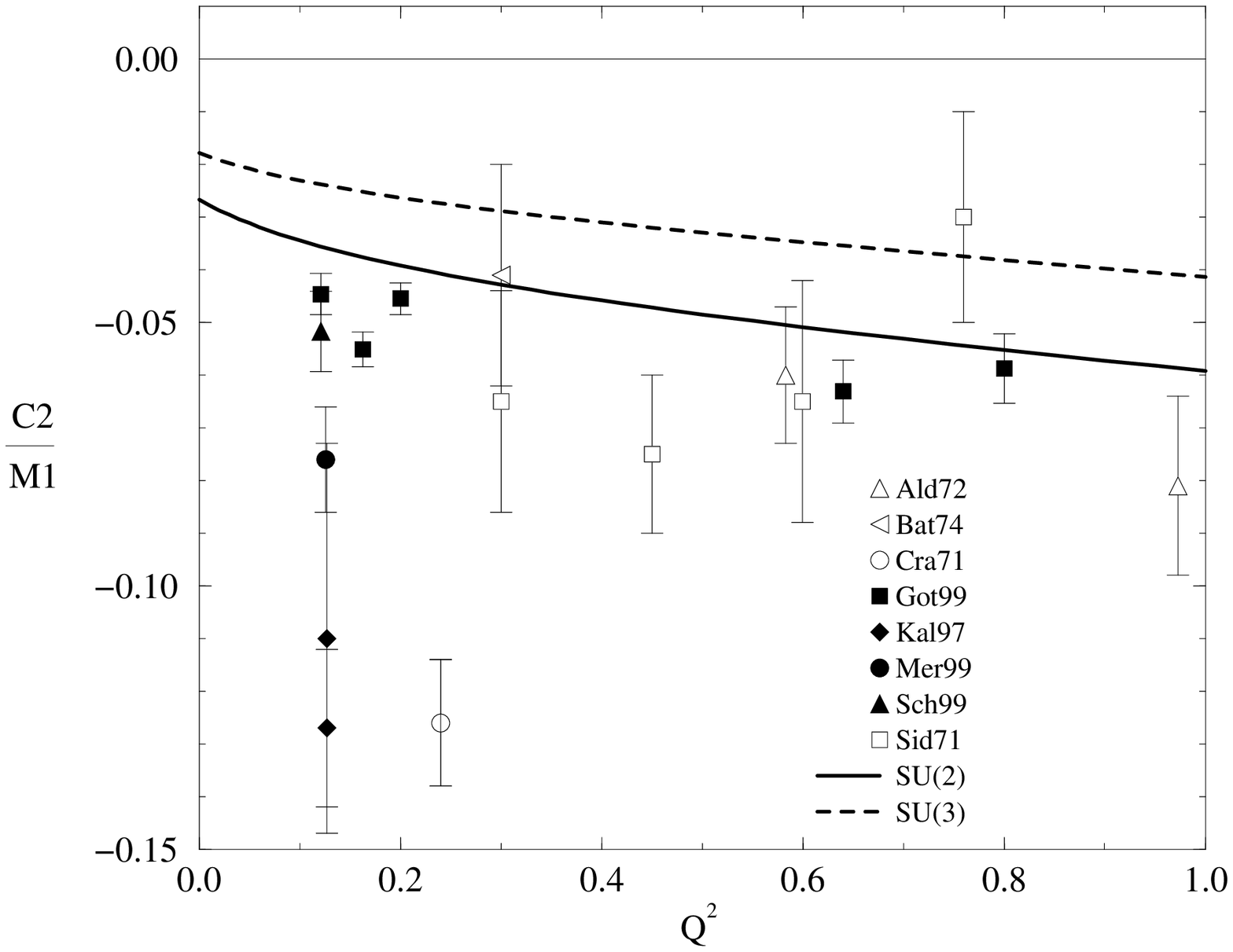}
\caption[]{\protect\small The ratio C2/M1 calculated in the CQSM, 
   in flavor SU(2) and SU(3), 
   for a constituent quark mass of 420 MeV 
   represented as a function of $Q^2$ in GeV$^2$. 
   Older experimental data (open symbols) 
   is taken from references~\cite{old_data1,old_data2} and  more recent data
   (filled symbols) is from~\cite{new_data,Got99}.  }    
\protect\label{fig:c2m1}
\end{figure}

\begin{figure}
\hspace{2cm}
\epsfxsize=10.5cm
\epsffile{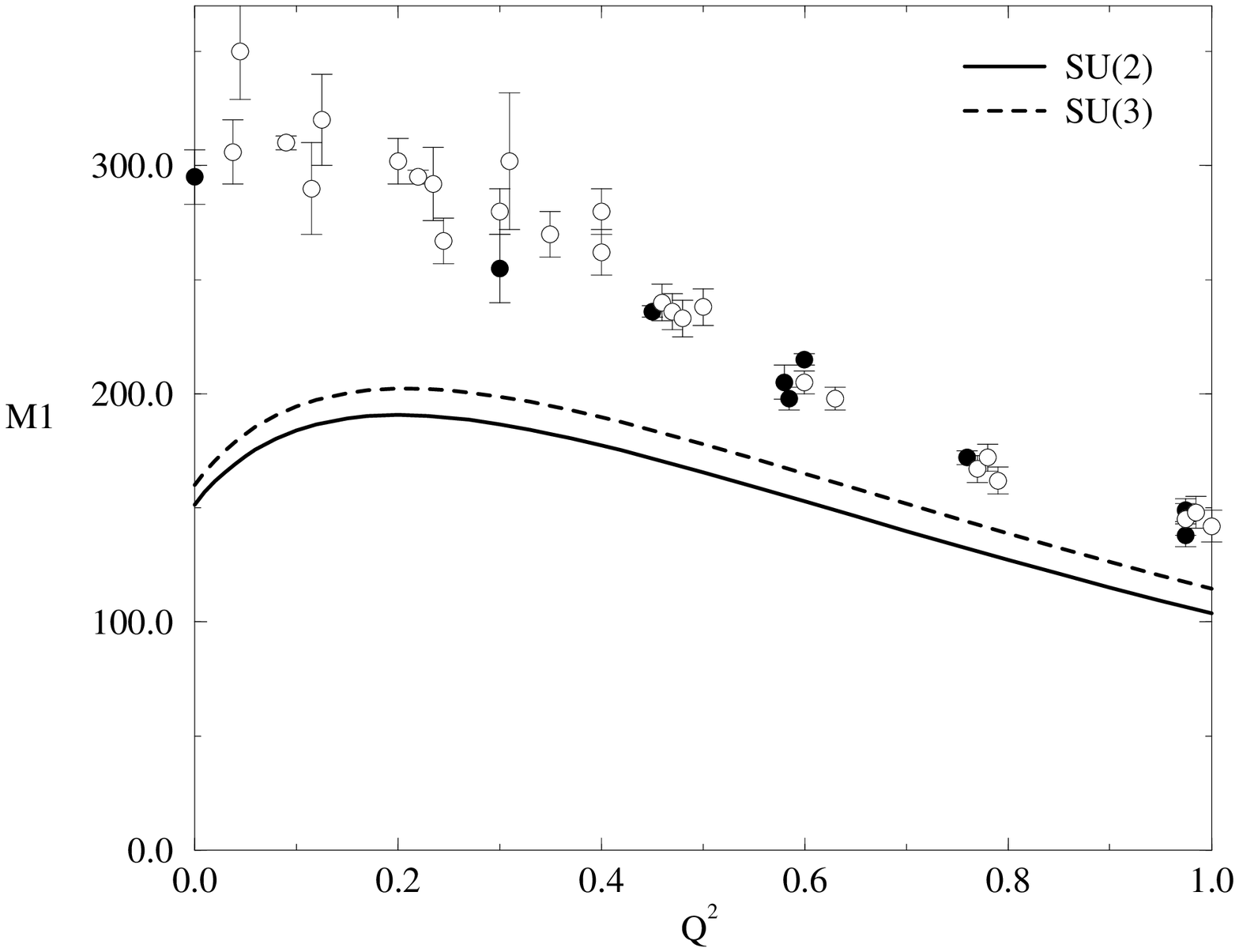}
\caption[]{\protect\small The magnetic 
   dipole amplitude M1${\scriptstyle =-\left[3A_{3/2}
   +\sqrt{3}A_{1/2} \right]/(2\sqrt{3})}$ calculated in 
   the CQSM, in flavor SU(2) and SU(3), 
   for a constituent quark mass of 420 MeV
   represented, in units of $10^{-3}$ GeV$^{-1/2}$, 
   as a function of $Q^2$ (in GeV$^2$). 
   The experimental data   
   is taken from  references~\cite{old_data2} ($\circ$) 
   and~ \cite{old_data3} ($\bullet$).  }  
\protect\label{fig:m1}
\end{figure}

The ratios  are related to the multipoles 
(\ref{eq:mm1}-\ref{eq:mc2}) through  
\begin{equation} {E2\over M1} = {1\over 3}{A_{1/2}(E2)\over A_{1/2}(M1)}
      = {1\over 3} {{\cal M}^{E2} \over {\cal M}^{M1} }  \end{equation} 
and
\begin{equation} {C2\over M1}={1 \over 2}{S_{1/2}(C2) \over A_{1/2}(M1)}
    ={1 \over 2\sqrt{2}} {{\cal M}^{C2} \over {\cal M}^{M1} }\, .    \end{equation}
Our results for them are presented in 
Fig.\ref{fig:e2m1} and in Fig.\ref{fig:c2m1}.
A first comparison allowed by these figures 
with the available experimental data,
allows us to conclude that the negative signs obtained for these 
two ratios are in agreement with the experimental data obtained in the
last few years. 
The multipole amplitude M1 is shown in Fig.~\ref{fig:m1} and compared
with experimental data~\cite{old_data2,old_data3}.
It underestimates the data and the situation does not improve
by considering the SU(3) case. This can be traced back to the model
 since it
also underestimates other magnetic-type observables. Actually this feature
is common to almost all of the hedgehog-type chiral soliton models and
apparently the present approach does not provide an exception for this
observable ~\cite{Kim96}.


For the ratio E2/M1, we obtain values, at the photon point, 
of $-2.1$~\% and  $-1.4$~\%, in SU(2) and SU(3), respectively.
A comparison with the most recent photoproduction data~\cite{photon_point}
in particular with the value  $-2.5\pm .5$~\%, noted by the 
Particle Data Group, 
reveals that our results are smaller, but that the SU(2) value still falls
within this estimate. Our prediction for finite momentum transfers yields
values which , starting at ${Q^2}=0$, tend momotonically to zero and vanish
basically for ${Q^2} \ge 0.6$ GeV$^2$. The predictions are in this
range qualitatively in
 agreement with the data in Ref.\cite{Got99}. The
change to positive E2/M1-values at $Q^2=0.8$ GeV$^2$ found by Gothe et
al.~\cite{Got99} is not reproduced in the present calculations.

For a direct comparison of our numbers with the above data we must take into
account that in our formalism the delta is a stable state, which does not
decay into nucleon and pion (unless the formalism is developed further which
is not done yet). Hence the model allows to calculate the real parts of the
transition amplitudes, but not the imaginary ones. Furthermore the extraction
of the  resonance contribution from the experimental data is not so easely
performed and is still a matter of debate. The difficulties are related,
among
 others, to the background contributions originating from
 the Born
term. This question of separating the contributions to the 
 amplitudes has
been adressed by several authors~\cite{Dav99} and in some
 cases the
non-resonance contributions were found to be large.  An unitary
 ambiguity is
known to exist in the case of the E2/M1  ratio~\cite{Wil98},
 related mainly
to the models necessary to extract the resonance contribution.
 A status
report concerning the latest determinations of E2/M1 can be found in
Ref.~\cite{Wor98}. Speed-plot techniques \cite{Han96} result in
E2/M1-ratios of about -3.5\%, which are larger than those quoted
 above. 
 
Concerning the comparison with other models, values for the ratio E2/M1 
 at
the photon point, obtained in the context of electroproduction
 studies, range
from $-0.2$~\% in a relativized quark 
 model~\cite{Cap90} to $-3.5$~\% ($Re$
E2/M1)
 in the context of heavy baryon chiral perturbation 
theory~\cite{Gel98}, passing
 through $-1.8$~\% in the chiral chromodielectric
model and $-1.9$~\%
 in the linear sigma model~\cite{Fio96}, $-2.3$~\% in the
skyrme 
 model~\cite{Wal97} and $-2.4$~\% ($Re$ E2/M1) in the chiral
bag model~\cite{Lu97}.  
Although the comparison between models may help to understand the
physical reasons for the observed ratios, it is also necessary
to take into account that different ingredients are involved in the
different model calculations above. 
The models more suitable for a comparison
with the CQSM are the constituent quark model and the Skyrme model,
between which the CQSM is supposed to interpolate in the limits
of small and large soliton sizes, respectively. 
Indeed, we find that our results in flavour  SU(2) at the photon 
point are between
the value of $-0.2$~\%~\cite{Cap90} for the constituent 
quark model and the values in the range
$-2.6$~\% to $-4.9$~\%~\cite{Wir87} obtained in the Skyrme model.
One has also to add however that the nonrelativistic quark model can
accomodate higher values, up to $-2$~\%, for the ratio E2/M1~\cite{Dre84} if the 
effects of basis truncation are taken with care. 
In which regards SU(3), our value  $-1.4$~\% for E2/M1 at the photon point
is smaller than the results for this ratio obtained in 
hyperon decays within the Skyrme model~\cite{skyrme},
which fall in the interval $-2.06$ to $-3.14$~\% for different model approaches.

It is interesting to note that in subsequent and more
refined calculations in these models the results approach an
intermediate common value, thus decreasing the difference in the
predictions: $-2.3$~\%~\cite{Wal97} in the Skyrme model and 
$-3.5$~\%~\cite{Buc98} for the constituent quark model
 (in photoproduction).
The numerical results obtained in~\cite{Buc98} show 
also the importance of the pionic degrees of freedom, both when compared 
with previous results in the constituent quark model and 
with the CQSM and the Skyrme models above, which already include 
such degrees of freedom from the very beginning, to different extents, though.
The role of the meson degrees of freedom may 
explanain the lower value obtained in SU(3) as compared to SU(2). It may be
caused by the poor description of the kaon cloud since the embedding of SU(2)
in SU(3) imposes a pion tail for the soliton giving the kaon too large an
importance.

For the ratio C2/M1, a comparison made in the same 
spirit as above for E2/M1,
reveals that this ratio slightly underestimates the data above 
$Q^2=0.3$ GeV$^2$, where other models~\cite{Fio96,Wal97}
obtain a better agreement. Nevertheless, our results are closer
to experiment than the quark model results~\cite{quark_model,Cap90}.   
As for values of $Q^2$ between $0$ and $0.3$ GeV$^2$, experimentally
the situation is not clear. While some old~\cite{old_data1}
and more recent~\cite{new_data} data seem to show a 
peaking structure around $0.15$~GeV$^2$, more recent 
results~\cite{Got99} still show a clear decrease of C2/M1 however
less pronounced with values similar to those observed for $Q^2>0.3$ Gev$^2$.
As far as we known, no model predicts such a peaking structure. 
Instead, our
results show a smooth growth  of the magnitude of the ratio C2/M1 with $Q^2$.
This behaviour is found within most of the other 
models in the range of $Q^2$ smaller than $1$~GeV$^2$ studied.
Our results are closer to experimental results, together 
with ~\cite{Fio96,Wal97}, when compared with the remaining models.

\section{Summary and Conclusions}

The photo- and electroproduction of the $\Delta(1232)$ have been 
investigated in the chiral quark-soliton model through the 
computed transition ratios, E2/M1 and C2/M1.
The three (four) parameters of the model
in SU(2) (SU(3)) were adjusted to the pion decay constant, the pion mass
(kaon mass) and from a general fit to nucleon
properties. No parametrization adjustment to nucleon-delta transitions was
considered.  Both ratios E2/M1 as well of C2/M1 are found to be
negative for all momentum transfers, as indicated by experiment. The value of
E2/M1 at $Q^2=0$ underestimates the most recent experimental
points by \mbox{30 \%} if one compares the numbers directly. 
This is the accuracy of the calculations also for finite momentum transfers. 
Strong fluctuations of
C2/M1 at small momentum transfers, as found in some experiments, are not seen
in the present approach. Generally the SU(3) calculations do not
improve the SU(2) results. 

\bigskip
%
%
\noindent {\large\bf Acknowledgements} 
\bigskip

The authors thank P.~Pobylitsa and M.~Polyakov for useful discussions. 
DU acknowledge the finantial support from PraxisXXI/BD/9300/96
and AS support from PraxisXXI/BD/15681/98 in the final stages of this work.
AS and MF acknowledge financial support from Inida Project \mbox{n432/DAAD.}
This work was also supported by PRAXIS/PCEX/P/FIS/6/96 
(Lisboa) and partially by the
DFG (Schwerpunktprogramm) and the COSY-Project Juelich.  

\appendix
\section{The Proper-time Regularization}
The proper-time regularization is based on the relation,
here applied to the real part of the 
fermionic determinant, 
\begin{eqnarray}
\lefteqn{ \left. -Re\left[ Tr \log  D(U) \right]
  +Re\left[ Tr \log  D(U\!=\!1) \right] \right|_{reg}  }
  \mbox{\hspace{3cm}}\nonumber\\
   & & = {1\over 2} {\rm Tr}
  \int_0^\infty \frac{du}{u}\phi \left(u,\Lambda\right)
  \left[ e^{-u D^\dagger(U) D(U)}-
  e^{-u D^\dagger(1) D(1)} \right]
\label{eq:reg} \end{eqnarray}
where the arbitrary function $\phi(u,1/\Lambda)$ 
is chosen as to make the integral finite: it has the form
$\theta(u-1/\Lambda^2)$ in SU(2) and (\ref{eq:phisu3}) in SU(3).

In order to obtain a finite expression for the sea 
contribution (\ref{eq:sea}), it is necessary to regularize~\cite{Chr95,Kim96a}
\begin{equation}   
 {\rm Tr} \log [D(U)-\epsilon O]={\rm Tr} \log [D(U_c)+i\Omega-\epsilon O] ,
\end{equation}
in terms of which the functional trace in (\ref{eq:sea})
can be rewritten through
\begin{equation}   
 {\rm Tr} \left[O\langle 0,0|-D^{-1}(U)|0,0\rangle  \right]=
 {\delta\over\delta\epsilon(0)}{\rm Tr} \left[D(U)-\epsilon O
 \right|_{\epsilon =0}.
\end{equation}
considering $\epsilon$ real but the operator $O$ not necessarely 
hermitian.

Since, for the operators considered here, only the real part 
of ${\rm Tr}\log D_\epsilon(U)$ is divergent, it is enough to consider
(\ref{eq:reg}). Using
\begin{eqnarray}   
 e^{\hat{A}+\hat{B}}-e^{\hat{A}}&=&
  \int\limits_0^1d\alpha\:e^{\alpha\hat{A}}\hat{B}
  e^{(1-\alpha)\hat{B}} \nonumber\\
 &+&\int\limits_0^1 d\beta\int\limits_0^{1-\beta}
  d\alpha\:e^{\alpha\hat{A}}\hat{B}e^{\beta\hat{A}}
  \hat{B}e^{(1-\alpha-\beta)\hat{A}} + \cdots
\end{eqnarray}
and expanding up to first order in the angular velocity $\Omega$ 
(\ref{eq:omegaexp}) leads to an expression similar to (\ref{eq:densis})
for the valence part, with ${\cal V}^{(\Omega^0)}$ and ${\cal V}^{(\Omega^1)}$
substituted respectively by ${\cal S}^{(\Omega^0)}$ and ${\cal S}^{(\Omega^1)}$
given by
\begin{equation}   
 {\cal S}^{(\Omega^0)}(\mbox{\boldmath $x$})=\sum_n {\cal R}_1
 (\varepsilon_n,\eta)\phi^\dagger_n(\mbox{\boldmath $x$})\,A^\dagger O A\,
 \phi_n(\mbox{\boldmath $x$}) , 
\end{equation}
\begin{eqnarray}   
 {\cal S}^{(\Omega^1)}(\mbox{\boldmath $x$})&=&{1\over 4I_1}
 \sum_{m,n}\Bigl[ {\cal R}_2^{(+)}(\varepsilon_m,\varepsilon_n,\eta)
 J_a[\phi^\dagger_m(\mbox{\boldmath $x$})\,A^\dagger O A\,
 \phi_n(\mbox{\boldmath $x$})] \nonumber\\
 &&\mbox{\hspace{1cm}}+
 {\cal R}_2^{(-)}(\varepsilon_m,\varepsilon_n,\eta)
 [\phi^\dagger_m(\mbox{\boldmath $x$})\,A^\dagger O A\,
 \phi_n(\mbox{\boldmath $x$})]J_a \Bigr]
 \langle n|\tau_a|m\rangle , 
\end{eqnarray}
again in SU(2) for simplicity. The constant $\eta$ is defined by
$O^\dagger=\eta O$.

The regularization functions appearing in the final expressions for
the amplitudes are written in terms of the functions above according
to~\cite{Chr95,Kim96a}: 
\begin{equation}   
 R_{1}\left( \varepsilon_n \right) \equiv 
 R_{1}\left( \varepsilon_n ,-1\right) =
 -{\varepsilon _n\over \sqrt{\pi}} 
 \int_0^{\infty } \frac{du }{\sqrt{u } }
 \phi(u,\Lambda)e^{-u \varepsilon_n^2}\, , 
\end{equation}
\begin{eqnarray}  
\lefteqn{  R_{A}\left( \varepsilon_m,\varepsilon_n\right) =
 -{1\over2}\left[{\cal R}_2^{(+)}(\varepsilon_m,\varepsilon_n,-1)
 + {\cal R}_2^{(-)}(\varepsilon_m,\varepsilon_n,-1)\right]
 } \nonumber\\
 &&= -\frac{1}{2\sqrt{\pi }}
 \int_0^{\infty } \frac{du }{\sqrt{u } } \phi(u,\Lambda)
 \left(   \frac{\varepsilon_n e^{-u\varepsilon_n^2}+\varepsilon_m 
 e^{-u\varepsilon_m^2}}{\varepsilon_m^{{}}
 +\varepsilon_{n}^{{}}} - \frac{1}{u}
 \frac{e^{-u\varepsilon_n^2}-e^{-u\varepsilon_m^2}}
 {\varepsilon_m^2-\varepsilon_n^2} \right) \, ,       
\end{eqnarray}
\begin{eqnarray}  
\lefteqn{  R_B\left( \varepsilon_m,\varepsilon_n\right) =
 {1\over2}\left[{\cal R}_2^{(-)}(\varepsilon_m,\varepsilon_n,1)
 - {\cal R}_2^{(+)}(\varepsilon_m,\varepsilon_n,1)\right]
  } \nonumber\\
 && = -\frac{1}{2\pi }\int_0^{\infty } \!du 
  \phi(u,\Lambda) \int_{0}^{1}\!d\alpha 
  \frac{\left(1-\alpha \right) \varepsilon_m-\alpha\varepsilon_n}
 {\sqrt{\alpha \left(1-\alpha \right) }}\:
 e^{-u\left[\alpha \varepsilon_n^2+\left(1-\alpha \right) 
 \varepsilon_m^2 \right] /\Lambda^2} \, .
\end{eqnarray}

\end{document}